\documentclass[conference]{IEEEtran}
\ifCLASSINFOpdf
\else
\fi
\usepackage{graphicx}
\usepackage{tabularx,booktabs}
\usepackage{dingbat}
\usepackage{diagbox}
\usepackage{multirow} 
\usepackage[hyphens]{url}
\usepackage[hidelinks,breaklinks]{hyperref}
\usepackage{xcolor}
\usepackage{amsfonts, amsmath, amsthm, amssymb}
\usepackage{subcaption}
\hypersetup{breaklinks=true}
\usepackage{tikz}
\usepackage{textcomp}
\usepackage{lipsum}
\IEEEoverridecommandlockouts
\newcommand\copyrighttext{%
  \footnotesize \textcopyright 2023 IEEE.  Personal use of this material is permitted.  Permission from IEEE must be obtained for all other uses, in any current or future media, including reprinting/republishing this material for advertising or promotional purposes, creating new collective works, for resale or redistribution to servers or lists, or reuse of any copyrighted component of this work in other works.}
\newcommand\copyrightnotice{%
\begin{tikzpicture}[remember picture,overlay]
\node[anchor=south,yshift=10pt] at (current page.south) {\fbox{\parbox{\dimexpr\textwidth-\fboxsep-\fboxrule\relax}{\copyrighttext}}};
\end{tikzpicture}%
}
\usepackage{graphicx}
\graphicspath{{figures/}}

\begin{document}
%
\title{Pensieve 5G: Implementation of RL-based ABR Algorithm for UHD 4K/8K Content Delivery on Commercial 5G SA/NR-DC Network}

\author{
    \IEEEauthorblockN{Kasidis ARUNRUANGSIRILERT\IEEEauthorrefmark{1}, Bo WEI\IEEEauthorrefmark{1}, Hang SONG\IEEEauthorrefmark{2}, Jiro KATTO\IEEEauthorrefmark{1}}
    \IEEEauthorblockA{\IEEEauthorrefmark{1}Department of Computer Science and Communications Engineering, Waseda University, Tokyo, Japan}
    \IEEEauthorblockA{\IEEEauthorrefmark{2}Department of Transdisciplinary Science and Engineering, Tokyo Institute of Technology, Tokyo, Japan
    \\\{kasidis, weibo, katto\}@katto.comm.waseda.ac.jp, song.h.ac@m.titech.ac.jp}
}
%
\maketitle
\copyrightnotice
\begin{abstract}
While the rollout of the fifth-generation mobile network (5G) is underway across the globe with the intention to deliver 4K/8K UHD videos, Augmented Reality (AR), and Virtual Reality (VR) content to the mass amounts of users, the coverage and throughput are still one of the most significant issues, especially in the rural areas, where only 5G in the low-frequency band are being deployed. This called for a high-performance adaptive bitrate (ABR) algorithm that can maximize the user quality of experience given 5G network characteristics and data rate of UHD contents.

Recently, many of the newly proposed ABR techniques were machine-learning based. Among that, Pensieve is one of the state-of-the-art techniques, which utilized reinforcement-learning to generate an ABR algorithm based on observation of past decision performance. By incorporating the context of the 5G network and UHD content, Pensieve has been optimized into Pensieve 5G. New QoE metrics that more accurately represent the QoE of UHD video streaming on the different types of devices were proposed and used to evaluate Pensieve 5G against other ABR techniques including the original Pensieve. The results from the simulation based on the real 5G Standalone (SA) network throughput shows that Pensieve 5G outperforms both conventional algorithms and Pensieve with the average QoE improvement of 8.8\% and 14.2\%, respectively. Additionally, Pensieve 5G also performed well on the commercial 5G NR-NR Dual Connectivity (NR-DC) Network, despite the training being done solely using the data from the 5G Standalone (SA) network.
\end{abstract}

\begin{IEEEkeywords}
5G Network, 5G Standalone, Adaptive Bitrate, UHD Video, QoE
\end{IEEEkeywords}


%
\IEEEpeerreviewmaketitle

\section{Introduction}
In the early 2020s, the deployment of 5G networks started around the world with goals to enable the new application of mobile networks for Ultra High Definition (UHD) video streaming, Virtual Reality (VR)/Augmented Reality (AR) content delivery, Smart City, Industrial Automation, Self Driving Vehicle, etc. There are mainly two types of 5G deployment that are being deployed commercially: 5G Non-Standalone (NSA), and 5G Standalone (SA). In the early stage of 5G deployment, mobile network operators will utilize the 5G NSA architecture and deploy only 5G base station (gNodeB) without deploying 5G Core. This allows the network operators to rollout 5G very quickly and lower the cost of initial deployment. While this type of deployment allows fast roll-out during the initial stage of 5G deployment, the usage of 4G Long-Term Evolution (LTE) signaling from the legacy Evolved Packet Core (EPC) means that features such as network slicing and Ultra-Reliable Low Latency Communications (URLLC) are unavailable. The ultimate goal of 5G evolution is to migrate toward 5G SA architecture with the deployment of 5G Core, then switch to 5G Signaling, which supports new 5G features \cite{9069723}.

\begin{figure}[t!]
  \centering
  \includegraphics[width=0.45\textwidth]{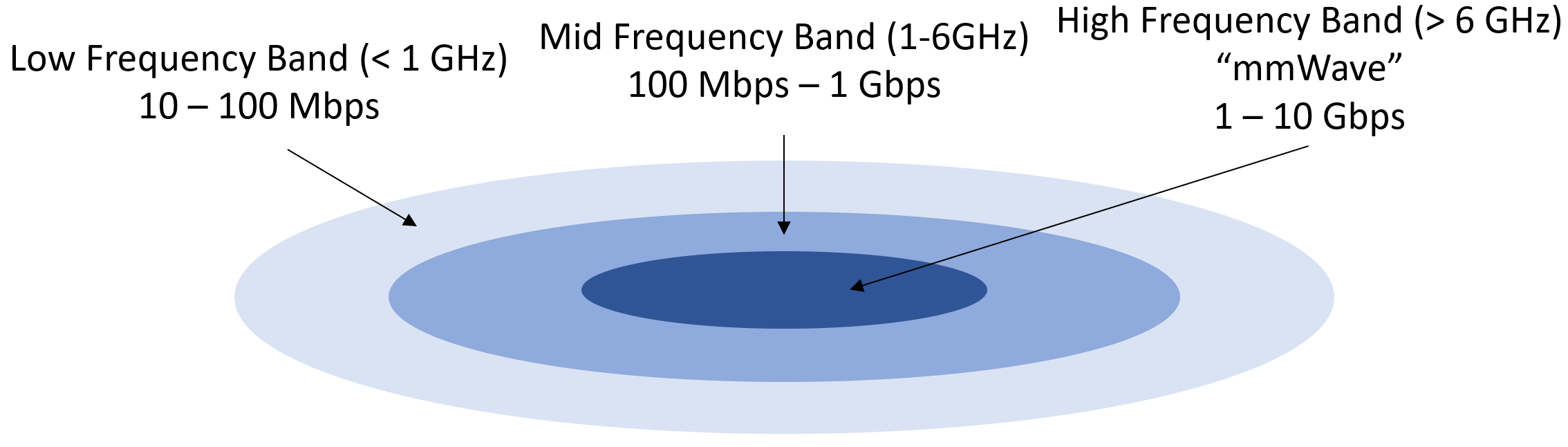}
  \caption{Coverage of Different 5G Frequency Bands}
  \label{fig:FreqBand}
\end{figure}

\begin{figure}[t!]
  \centering
  \includegraphics[width=0.39\textwidth]{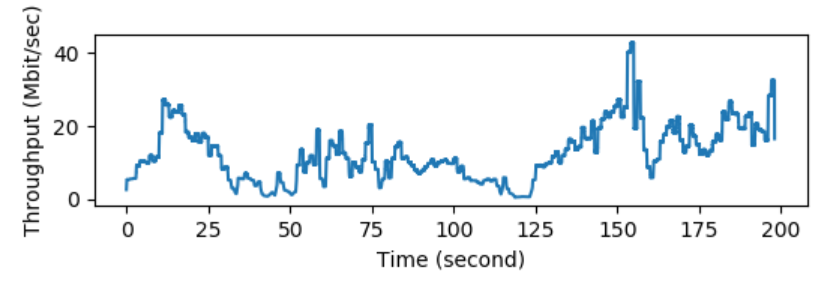}
  \setlength{\belowcaptionskip}{-15pt}
  \caption{Real-World Throughput of Low Frequency 5G Band in Suburban area in Saitama Prefecture, Japan}
  \label{fig:LowThpt}
\end{figure}

While 5G promises to provide up to 20 Gbps peak downlink throughput and 10 Gbps peak uplink throughput with low latency \cite{itu_m2083}, that speed is only achievable in the area with High Frequency Band (mmWave) service \cite{8886119}\cite{9045247}, which is known to have poor coverage and intended for the very high-density urban environment only. In the suburban and rural areas, the middle frequency band and low-frequency band are commonly used (see Figure \ref{fig:FreqBand}). The middle frequency band are capable of delivering throughput up to 1.7 Gbps in ideal condition with typical real-world throughput of around 100 Mbps to 1 Gbps \cite{9497810}, while the low-frequency band is usually deployed on the existing LTE frequency spectrum either by migrating toward 5G completely or by the use of Dynamic Spectrum Sharing (DSS). The low-frequency band only provides the basic service with typical throughput of lower than 100 Mbps. As networks migrate toward the SA architecture, low-frequency bands, formerly used for LTE networks, are heavily used to extend 5G coverage in suburban and rural areas. 5G deployed in the low-frequency band only provides a latency advantage over the existing LTE network while offering no throughput uplift at all. This means that the delivery of UHD videos and VR/AR content over 5G in suburban and rural areas can poses a significant challenge due to the limited throughput. Figure \ref{fig:LowThpt} illustrates the typical throughput of the 5G SA network on a train in the suburban area in Saitama Prefecture, Japan.
\begin{figure}[t!]
  \centering
  \includegraphics[width=0.4\textwidth]{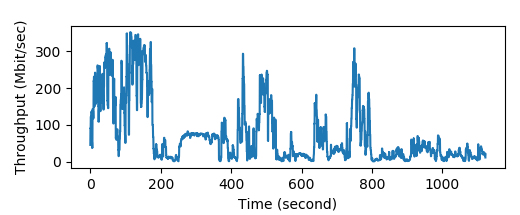}
  \setlength{\belowcaptionskip}{-25pt}
  \caption{SoftBank 5G SA Throughput on JR Yamanote Line in Tokyo, Japan}
  \label{fig:Handover}
\end{figure}

In the real-world situation, it's inevitable that the users will be moving between the service area of different frequency bands, which offer different maximum theoretical throughput. While high and middle-frequency bands are adequate for UHD video delivery, the low-frequency band might not be able to provide high enough throughput for smooth playback, resulting in bad Quality of Experience (QoE) due to rebuffering and stuttering. This may cause the user to give up the service causing a loss of business and profit. While HTTP adaptive streaming techniques such as MPEG-DASH have been evaluated to performed considerably good on 5G middle frequency band with Massive MIMO deployment \cite{9868877}, it can not keep up very well when the user equipment (UE) is being handover between different frequency bands with different maximum throughput because the throughput range became very broad as seen in Figure \ref{fig:Handover}. Therefore, more advanced ABR algorithm should be used to cope with the issue.

Due to the unpredictable nature of Mobile Network, conventional Adaptive Bitrate (ABR) algorithms that utilize a set of rules are unlikely to successfully deliver good QoE. Pensieve \cite{10.1145/3098822.3098843} has been proved to be one of the most successful implementations of Reinforcement Learning (RL) in ABR streaming applications. However, the original work was optimized for 3G Universal Mobile Telecommunications System (UMTS) and 4G LTE network, which has different characteristics than 5G New Radio (NR). Therefore, in this paper, Pensieve will be optimized for 5G NR and evolved to "Pensieve 5G." New QoE metrics based on recent works on Quality of Experience on different types of devices will be introduced, then will be used to evaluate and compared Pensieve 5G to other adaptation algorithms. Finally, Pensieve 5G will be evaluated on the world's first 5G NR Dual Connectivity (5G NR-DC) deployment \cite{qualcomm_2022}, allowing the combination of middle frequency 5G band (sub-6 GHz) to high frequency 5G (mmWave) band, on NTT DoCoMo, the largest mobile network operator in Japan. The rest of the paper is organized as follows: Section II will provide some of the related works, Section III describes how the experiment is being configured and introduces newly proposed QoE metrics, Section IV presents the experiment results, and the conclusion and future work should be discussed in Section V.
\section{Related Work}

Conventional ABR algorithms work by one of the two following principles: estimating the network throughput (Rate-based), or monitoring playback buffer status (Buffer-based). A rate-based algorithm such as FESTIVE \cite{10.1145/2413176.2413189} works by monitoring past chunk download activities or utilize other techniques such as TRUST \cite{8647390} in order to make a prediction about the network throughput, then attempt to fetch the best quality chunk based on the prediction. While Buffer-based algorithms such as BOLA \cite{9110784}, and FRAB \cite{9500784} attempt to keep their buffer level at the target value. Latest works such as BOLA have the ability to stop the chunk download, then switch to the lower quality one if the download time is deemed excessive. These methods may work well with legacy 3G UMTS or 4G LTE networks due to the fixed maximum carrier bandwidth of 5 MHz and 20 MHz, respectively. However, the introduction of carrier aggregation (CA) \cite{5493902} in LTE-Advanced (LTE-A) resulted in highly unpredictable throughput that varies by a huge margin from one base station to another as mobile network operators might not deploy all frequency bands on every base station.

Recently, Y. Yuan et al. proposed VSiM \cite{9796725} to improve QoE Fairness for Mobile Video Streaming. The system works by detecting the client's mobility profile, then triggering the server push mechanism to push multiple low-quality chunks to fill the client's buffer using QUIC protocol when the server detected that the client buffer is starving. While the system is quite robust against rebuffering and provides state-of-the-art performance in QoE fairness, mobility profile detection works by actively obtaining GPS location from clients, which may cause excessive battery drain and privacy concerns.

\section{Test Configurations}
\subsection{Throughput Data Collection} \label{subsectionA}
To get a good variety of training data, data from multiple network operators across two countries that provides commercial 5G SA service were obtained. This includes Advanced Info Service (AIS) Thailand, SoftBank Japan, and NTT DoCoMo Japan. Additionally, multiple types of User Equipment with a variety of network capabilities were used including Samsung Galaxy Z Flip3 5G (SM-F711B), Samsung Galaxy S22 Ultra (SC-52C), and ASUS Smartphone for Snapdragon Insiders (EXP21), which should provide generalized data that is not unique to one specific model of user equipment or network. To collect the data, mobile network testing tools called "Network Signal Guru" from Qtrun Technologies were used. Test cases were set up to download a large file via HTTP protocol for an infinite amount of time with the option to automatically restart upon connection loss turned on. The log file was imported into "AirScreen" software from the same company, then the application layer throughput with timestamp was exported into CSV format. The log file also contains RF signal parameters, which will be used in future works. In contrast to the original Pensieve, in which only the traces with average throughput from 0.2 Mbps to 6.0 Mbps were included, all traces regardless of minimum or maximum throughput were included in this work to represent the broad range of throughput possible in real-world 5G NR network.

To cover typical mobility and everyday use cases. The throughput collection was done when the UE was at stationary, moving at walking speed, traveling on cars, trains, and streetcars. The location of the data collection also varies from urban, suburban, and rural, which should provide a good representation of mobile network conditions from close to ideal conditions to worst-case scenarios. Additionally, UE was taken into the concert to obtain the data while the base station was heavily loaded by multiple users. Lastly, the data was obtained from the service area of the next generation 5G deployment, the 5G NR-DC on NTT DoCoMo Network.

\begin{figure*}[h!]
\centering
\begin{subfigure}{.33\textwidth}
  \centering
  \includegraphics[width=0.98\linewidth]{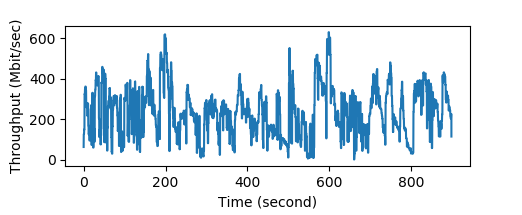}
  \caption{5G SA - Driving}
  \label{fig:Driving}
\end{subfigure}%
\begin{subfigure}{.33\textwidth}
  \centering
  \includegraphics[width=0.98\linewidth]{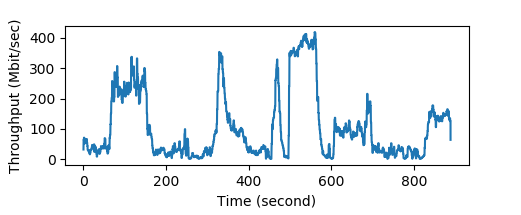}
  \caption{5G SA - Streetcar}
  \label{fig:StreetCar}
\end{subfigure}%
\begin{subfigure}{.33\textwidth}
  \centering
  \includegraphics[width=0.98\linewidth]{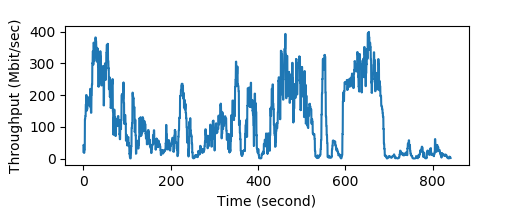}
  \caption{5G SA - Suburban Train}
  \label{fig:SuburbTrain}
\end{subfigure}
\begin{subfigure}{.33\textwidth}
  \centering
  \includegraphics[width=0.98\linewidth]{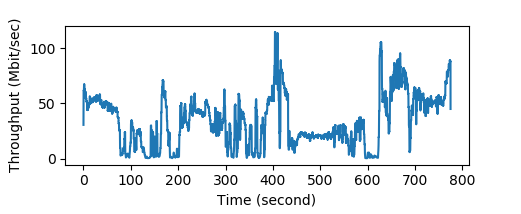}
  \caption{5G SA - Rural Train}
  \label{fig:RuralTrain}
\end{subfigure}%
\begin{subfigure}{.33\textwidth}
  \centering
  \includegraphics[width=0.95\linewidth]{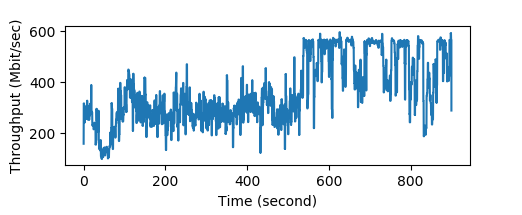}
  \caption{5G SA - Concert}
  \label{fig:Concert}
\end{subfigure}%
\begin{subfigure}{.33\textwidth}
  \centering
  \includegraphics[width=0.98\linewidth]{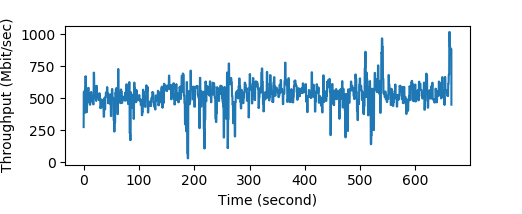}
  \caption{5G SA NR-DC - Walking}
  \label{fig:NRDC}
\end{subfigure}
\setlength{\belowcaptionskip}{-15pt}
\caption{Summarize of 5G Throughput Characteristics in Various Scenarios}
\end{figure*}

\subsection{Bitrate, Buffer Length, and Encoding Setting} \label{subsectionB}
In the original Pensive \cite{10.1145/3098822.3098843}, the author used the buffer length of 60 seconds with six representations of the video with the data rate from 300 kbps to 4.3 Mbps, which covers the resolution from 240p to 1080p. The chunk length of 4 seconds was used. The encoding configuration and buffer length was obtained from YouTube using youtube-dl. The bitrate of AV1 encoding of eight 8K videos was obtained, then the median value was obtained. The average was not used because some of the YouTube videos were granted significantly higher bitrate than the typical bitrate and became an outlier. Additionally, by selecting the video resolution, pausing the video, then using the "Stats for nerds" option on the YouTube playback page, the buffer length for each resolution can be obtained. Table \ref{tab:YouTubeConfig} shows the median bitrate, buffer length, and buffer size of YouTube.

\begin{table}[!tbp]
\caption{YouTube Median AV1 Bitrate, Buffer Length, and Buffer Size for each video resolution}
\centering
\label{tab:YouTubeConfig}
\resizebox{6.7cm}{!}{\begin{tabular}{@{}llll@{}}
\toprule
Resolution                      & Median  & Buffer & Buffer\\
 & Bitrate (kbps) & Length (s)&Size (MB)\\\midrule
144p                            & 83  & 120 & 1.3                                 \\
240p                            & 181 & 120 & 2.7                                               \\
360p                            & 397 & 120 & 6.1      \\
480p                            & 725 & 120 & 10.6                                                 \\
720p                            & 1861 & 120 & 28.8                              \\
1080p (Full HD)                           & 3372 & 120 & 51.3                                               \\
1440p                           & 8706 & 57.5 & 94.2            \\
2160p (UHD 4K)                           & 17941 & 30.6 & 97.5                                                    \\
2880p                           & 13873 & 56.7 & 103.9                                                    \\
4320p (UHD 8K)                           & 37087 & 23.8 & 101.6       \\ 
\bottomrule
\end{tabular}}
\vspace{-6mm}
\end{table}

From YouTube configuration, it has been found that the web browser has a media buffer at a maximum of around 100 MB, so the buffer length of 24 seconds was chosen. To avoid excessively large chunk size due to the high bitrate of 8K UHD video, the chunk size was reduced from 4 seconds to 2 seconds. Additionally, due to the lack of AV1 decoding capability on the experiment test bench, the video was encoded in VP9 codec with the same bitrate as the YouTube AV1, which should give a good representation of the AV1 payload. Finally, due to the 5K resolution on YouTube being experimental, the bitrate from YouTube was not used and a bitrate between 2160p (UHD 4K) and 4320p (UHD 8K) was chosen. The bitrate used for the experiment is summarized in Table \ref{tab:ExBitrate}. The source video was a gaming video with a resolution of 3840x2160p and a frame rate of 60 fps, the video was upscaled for the resolutions above 2160p (UHD 4K).
\begin{table}[!tbp]
\caption{Bitrate used for Experiment}
\centering
\label{tab:ExBitrate}
\resizebox{3.7cm}{!}{\begin{tabular}{@{}ll@{}}
\toprule
Resolution                      & Bitrate (kbps)  \\\midrule
144p                            & 100                                 \\
240p                            & 300                                               \\
360p                            & 500     \\
480p                            & 1000                                                \\
720p                            & 2000                            \\
1080p (Full HD)                 & 4000                                               \\
1440p                           & 8000           \\
2160p (UHD 4K)                  & 18000                                                    \\
2880p                           & 28000                                                    \\
4320p (UHD 8K)                  & 37500       \\ 
\bottomrule
\end{tabular}}
\vspace{-6mm}
\end{table}

\subsection{QoE Metrics}
The original works utilize the general QoE metric from MPC \cite{10.1145/1878022.1878030}, which can be defined as
\begin{equation}
QoE = \sum_{n=1}^{N} q(R_n)-\mu\sum_{n=1}^{N} T_n - \sum_{n=1}^{N-1}\mid q(R_{n+1})-q(R_n)\mid
\end{equation}

Given video with \begin{math}N\end{math} chunks, \begin{math}R_n\end{math} represent the bitrate for the \begin{math}chunk_n\end{math} and \begin{math}q(R_n)\end{math} represents the QoE perceived by the user given the bitrate. \begin{math}T_n\end{math} is the rebuffering time in seconds used to penalize the QoE by the factor of \begin{math}\mu\end{math} for the amount of time user need to wait for rebuffering. Finally, the change in video quality resulted in reduced smoothness and lowered the QoE is penalized by the final terms.

However, a study \cite{8019297} shows that the negative MOS impact only occurs when the resolution is being lowered from the higher quality representation to the lower one. Therefore, QoE should not be penalized for the increase in video quality. Therefore, the QoE metric should be defined as follows
\begin{equation} \label{eq:2}
QoE = \sum_{n=1}^{N} q(R_n)-\mu\sum_{n=1}^{N} T_n - S\sum_{n=1}^{N-1}(q(R_n)-q(R_{n+1}))
\end{equation}
Where S is defined as
\begin{equation} 
  S =
  \begin{cases}
    1 & \text{if $q(R_{n+1}) < q(R_n)$} \\
    0 & \text{otherwise}
  \end{cases}
\end{equation}

Six choices of \begin{math}q(R_n)\end{math} are considered, the first two are from the original work, while the latter four are newly purposed based on recent studies:
\begin{enumerate}
  \item \begin{math}QoE_{lin}: q(R_n)=R_n\end{math}: Used by MPC \cite{10.1145/2785956.2787486}.
  \item \begin{math}QoE_{log}: q(R_n)=log(R/R_{min})\end{math}: Used by BOLA \cite{9110784}, this metric assumes that the quality improvement became less perceivable at a higher bitrate.
  \item \begin{math}QoE_{HD}: 50(Resolution/4320)\end{math}: This assigns the score out of 50 by using the vertical resolution of the current chunk proportional to the vertical resolution of the best quality representation.
    \item \begin{math}QoE_{SMARTPHONE}: \end{math} QoE of watching the UHD video on smartphone out of 50, be taken into account the study \cite{8122249}, which results shows that QoE improvement on the smartphone in Non-VR environment is negligible beyond 1080p due to the limited resolution and small display size.
    \item \begin{math}QoE_{TV}: \end{math} QoE of watching the UHD video on an 8K Television set out of 50, be taken into account studies \cite{9506605}\cite{9524830}, which shows that the improvement in QoE of resolution beyond 2160p at typical viewing distance is minor.
    \item \begin{math}QoE_{VR}: \end{math} QoE of watching VR content a VR headset out of 50, be taken into account the study \cite{8122249}, which shows that the improvement in resolution is more perceivable than when viewing on smartphone likely due to the very close viewing distance.
\end{enumerate}

\begin{table}[!tbp]
\caption{QoE metrics used in evaluation. Each metric is used in Equation \ref{eq:2}}
\centering
\label{tab:QoEMetric}
\resizebox{8.7cm}{!}{\begin{tabular}{@{}lll@{}}
\toprule
QoE Metric                      & \begin{math}q(R)\end{math} &\begin{math}\mu\end{math} \\\midrule
\begin{math}QoE_{lin}\end{math}& \begin{math}R_n\end{math} & 37.5\\
\begin{math}QoE_{log}\end{math}& \begin{math}log(R/R_{min})\end{math} & 5.93\\
\begin{math}QoE_{HD}\end{math}& \begin{math}50(Resolution/4320)\end{math} & 24\\
\begin{math}QoE_{SMARTPHONE}\end{math}&\begin{math}144p	\rightarrow1,240p\rightarrow10,360p\rightarrow25,480p\rightarrow35,\end{math} & 25\\
&\begin{math}720p	\rightarrow42,1080p\rightarrow45,1440p\rightarrow47,\end{math} & \\
&\begin{math}2160p	\rightarrow48,2880p\rightarrow49,4320p\rightarrow50\end{math} & \\
\begin{math}QoE_{TV}\end{math}&\begin{math}144p	\rightarrow1,240p\rightarrow8,360p\rightarrow18,480p\rightarrow24,\end{math} & 45\\
&\begin{math}720p	\rightarrow30,1080p\rightarrow35,1440p\rightarrow42,\end{math} & \\
&\begin{math}2160p	\rightarrow46,2880p\rightarrow48,4320p\rightarrow50\end{math} & \\
\begin{math}QoE_{VR}\end{math}&\begin{math}144p	\rightarrow1,240p\rightarrow6,360p\rightarrow14,480p\rightarrow18,\end{math} & 50\\
&\begin{math}720p	\rightarrow25,1080p\rightarrow32,1440p\rightarrow38,\end{math} & \\
&\begin{math}2160p	\rightarrow42,2880p\rightarrow46,4320p\rightarrow50\end{math} & \\

\bottomrule
\end{tabular}}
\vspace{-6mm}
\end{table}

\subsection{Proposed Method}
Due to the increased complexity resulting from increasing the number of representations in MPEG-DASH from six to ten, the number of neurons and filters has been increased from 128 to 320, and the learning rate of the actor-network was reduced from \begin{math}1.0\times10^-4\end{math} to \begin{math}5.0\times10^-5\end{math}. Additionally, the reward metric used during training was changed from \begin{math}QoE_{lin}\end{math} used in original work to \begin{math}QoE_{HD}\end{math} with \begin{math}\mu\end{math} set to 80 instead of 24 during training to prevent the network from preferring to rebuffer high resolution representation over choosing the lower resolution one.

As for Pensieve's simulator, which is used to train the Pensive 5G neural network, the configuration was adjusted so that in case the buffer has filled up, the simulator will pause the video chunk download request for 2000ms instead of the original configuration of 500ms. Unlike legacy 4G LTE or 3G UMTS networks, the peak throughput of 5G NR is very high in the area with middle and high-frequency band service, which can easily cause the buffer to overflow. As one chunk of the video is 2000ms, this change prevents buffer overflow by ensuring that one chunk is played back and pruned from the buffer before adding a new chunk to the buffer. 

As for the training data, the original work used the throughput trace based on the 3G High-Speed Downlink Packet Access (HSDPA) network and filter out throughput data outside the range of 0.2 to 6.0 Mbps, which does not give a good representation of modern cellular network because the 3G HSDPA network has a fixed channel bandwidth and maximum throughput 5 MHz and 14.4 Mbps, respectively. However, modern LTE networks may have channel bandwidth between 1.4 MHz and 20 MHz, while 5G networks utilize channel bandwidth in the range between 5 MHz and 100 MHz. Therefore, the training data was replaced with throughput data on the 5G Standalone (SA) network, as mentioned in section \ref{subsectionA}, from various scenarios from high-throughput mid-band 5G Sub-6 Time Division Duplex (TDD) network in the urban area to fringe area coverage of low-band 5G Sub-6 Frequency Division Duplex (FDD) network in the rural area. Additionally, from all the training data, about 10\% of throughput data from the legacy 4G LTE network was also included because of the similarity in throughput characteristics between 4G LTE and 5G Sub-6 FDD network. Some throughput traces also include total loss of service to simulate challenging coverage areas of cellular networks. No throughput filtering was done to the data as the operation of the ABR algorithm in the challenging condition is one of the main objectives.

\begin{figure*}[h!]
  \centering
  \includegraphics[width=0.9\textwidth]{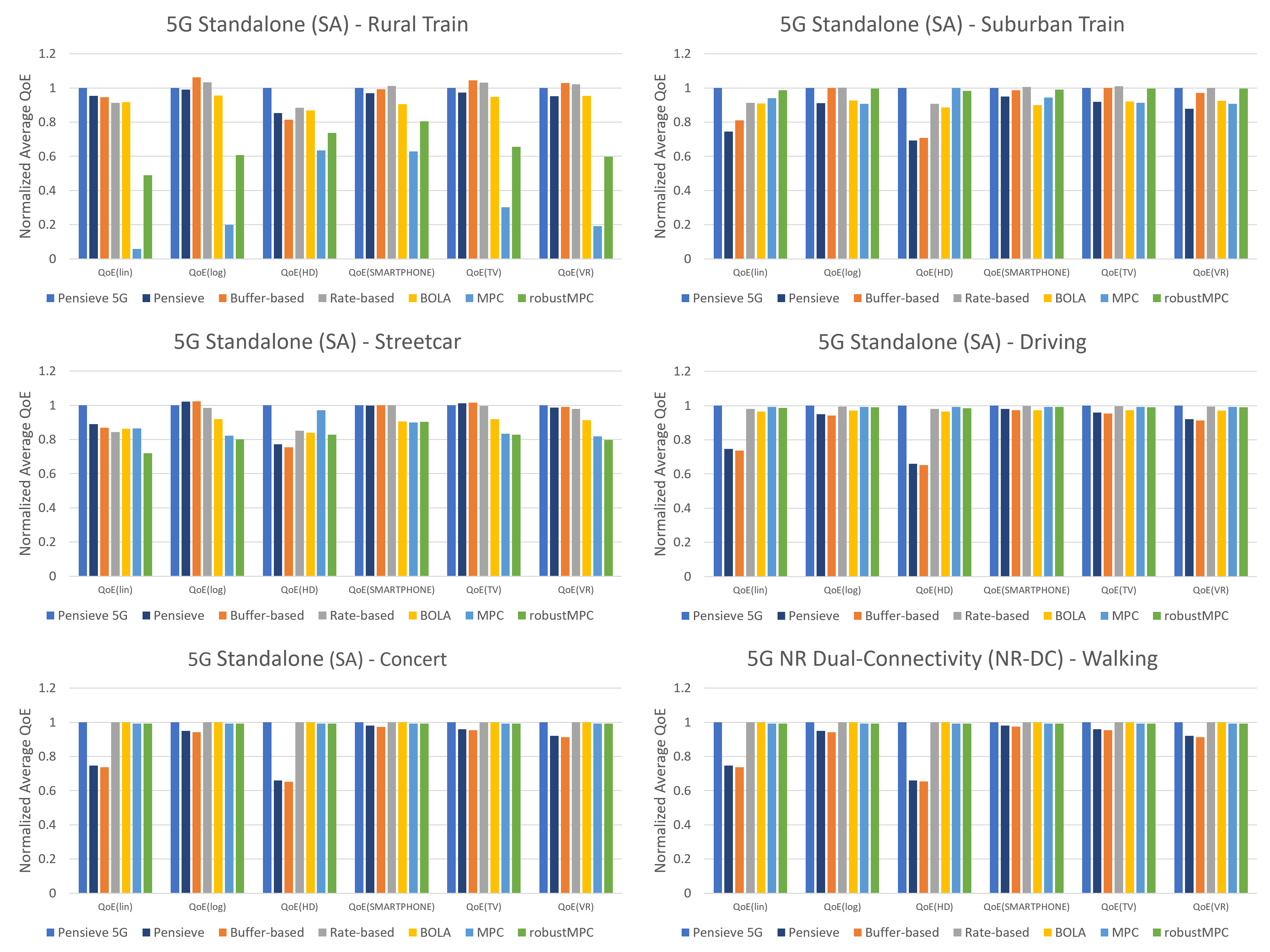}
  \setlength{\belowcaptionskip}{-15pt}
  \caption{Comparison of various ABR algorithms on various scenarios on 5G Standalone (SA) and 5G NR Dual-Connectivity (NR-DC) networks by using various QoE metrics. The results has been normalized against Pensieve 5G results.}
  \label{fig:Result}
\end{figure*}
\subsection{Evaluation}

To evaluate the performance of ABR algorithms, five 5G Standalone (SA) scenarios and one 5G NR-DC scenario were considered. 15-minute long throughput information was randomly chosen from each scenario, then the throughput data was converted to be compatible with Mahimahi \cite{10.5555/2813767.2813798} network emulation tool. A 13-minute long video encoded with the parameters in section \ref{subsectionB} was played back and the results were collected for each test case. To obtain the result for the original Pensieve, the training data was replaced with the same dataset as Pensieve 5G, but the training parameters were remain at the original configuration. The original Pensieve's simulator was used to train Pensieve, while Pensieve 5G was trained using the newly proposed configuration. Both Pensieve and Pensive 5G were trained for 120,000 epochs, then the best model was used for the evaluation.

The throughput characteristics of each scenario can be seen in Figure \ref{fig:Driving}, \ref{fig:StreetCar}, \ref{fig:SuburbTrain}, \ref{fig:RuralTrain}, \ref{fig:Concert}, and \ref{fig:NRDC}. Pensieve 5G, Pensieve \cite{10.1145/3098822.3098843}, Buffer-Based (BB) \cite{10.1145/2619239.2626296}, Rate-Based (RB), BOLA \cite{9110784}, MPC \cite{10.1145/2785956.2787486}, and robustMPC \cite{10.1145/2785956.2787486} were evaluated in this work.

\section{Experiment Results}
From the experiment, when comparing Pensieve 5G to conventional ABR algorithms. Pensieve 5G performed well overall across all considered case scenarios. The main focus of the discussion will be based on \begin{math}QoE_{HD}\end{math}, which reflects the raw performance, and \begin{math}QoE_{SMARTPHONE}\end{math}, which is the most likely use cases of typical users. In the cases with good middle-frequency band coverage such as streetcars, driving, and concert cases, when compared to other ABR algorithms, Pensieve 5G managed to deliver an average improvement of 4.8\% and 3.0\%, for \begin{math}QoE_{HD}\end{math} and \begin{math}QoE_{SMARTPHONE}\end{math}, respectively. When comparing to the original Pensieve, which did not design for large throughput range of 5G, Pensieve 5G yield 44.2\% better performance in \begin{math}QoE_{HD}\end{math}, but only providing 1.5\% improvement when considering \begin{math}QoE_{SMARTPHONE}\end{math}. It's worth mentioning that, in the cases with good connection quality, the original Pensieve and Buffer-Based (BB) ABR significantly underutilized the available bandwidth resulting in suboptimal performance. The 5G NR-DC case follows this trend, where all ABR algorithms performed well except Pensieve and BB, which underutilized the available bandwidth, resulting in Pensieve 5G performing an average of 17.5\% better than Pensieve across all QoE metrics on 5G NR-DC.

When considering the less ideal cases like Suburban Train and Rural Train, the Pensieve 5G yield an average of 14.8\% and 12.0\% improvement in \begin{math}QoE_{HD}\end{math} and \begin{math}QoE_{SMARTPHONE}\end{math}, respectively, over conventional ABR techniques. Additionally, when considering the original Pensieve, Pensieve 5G provides 31.0\% better \begin{math}QoE_{HD}\end{math} and 4.2\% uplift in \begin{math}QoE_{HD}\end{math} and \begin{math}QoE_{SMARTPHONE}\end{math}. Overall, Pensieve 5G provides an average of 8.8\% and 14.2\% better QoE across all metrics against conventional ABR techniques and Pensieve, respectively, demonstrating the superior performance suitable for the 5G network environment.

\section{Conclusions and Future Work}
In this paper, new QoE metrics for streaming of 4K/8K UHD contents on various devices were introduced to reflect the recent works and newly developed use cases. Then, the new configurations and training dataset based on a 5G network environment were proposed for Pensieve, a reinforcement-learning-based ABR technique, which developed into Pensieve 5G. The newly proposed QoE metrics as well as some from the original works were used to evaluate Pensieve 5G against the original Pensieve and other conventional ABR techniques. While Pensieve 5G did not outperform other ABR algorithms in all scenarios like the original work, it still delivers an average of 8.8\% better performance compared to conventional techniques and 14.2\% improvement over Pensieve, being able to take full advantage of the available network bandwidth to maximize the QoE in most cases including some of the challenging ones. Therefore, Pensieve 5G has the potential to become a useful technique to improve UHD 4K/8K video streaming QoE on the 5G NR network across various use cases and scenarios. As for future work, the incorporation of other RF parameters into the decision factor of the ABR algorithm should improve the performance of Pensieve 5G over the existing methods by a significant margin, especially in rural areas with challenging RF conditions and frequent complete loss of service.

\section*{Acknowledgement}
This work was supported in part by the Japan Society for the Promotion of Science KAKENHI (Grant No. 22K14299), in part by NICT (Grant No. 03801), Japan, in part by Waseda University-NICT Matching Research Project, and in part by the cooperation between Waseda University and Kioxia Corporation. Additionally, we would like to thank PEI Xiaohong of Qtrun Technologies for providing Network Signal Guru (NSG) and AirScreen, the cellular network drive test software used for result collection and analysis in this research. Finally, we thank Apidech Tearpaiboon for providing the 4K 60fps gaming video used in this research.




%
\Urlmuskip=0mu plus 1mu\relax
\bibliographystyle{IEEEtran}
\bibliography{IEEEabrv,reference}

\end{document}